\documentclass[sn-mathphys]{sn-jnl}

\jyear{2021}%

\theoremstyle{thmstyleone}%
\theoremstyle{thmstyletwo}%

\theoremstyle{thmstylethree}%

\usepackage{graphicx}
\usepackage{bm}
\usepackage{xcolor}

\begin{document}

\newcommand{\be}{\begin{equation}}
\newcommand{\ee}[1]{\label{#1}\end{equation}}
\newcommand{\bem}{\begin{eqnarray}}
\newcommand{\eem}[1]{\label{#1}\end{eqnarray}}
\newcommand{\eq}[1]{Eq.~(\ref{#1})}
\newcommand{\Eq}[1]{Equation~(\ref{#1})}
\newcommand{\ua}{\uparrow}
\newcommand{\da}{\downarrow}
\newcommand{\g}{\dagger}

\newcommand{\rc}[1]{\textcolor{red}{#1}}

\title[Comment on ``Square-well model...'']{Comment on ``Square-well model for superconducting pair-potential''}


\author*{\fnm{Edouard} \sur{Sonin}}\email{sonin@cc.huji.ac.il}

\affil{\orgdiv{Racah Institute  of Physics}, \orgname{Hebrew University of Jerusalem}, \orgaddress{\street{Givat Ram}, \city{Jerusalem}, \postcode{9190401}, 
 \country{Israel}}}


\abstract{The current-phase relation for a ballistic SNS junction in a 1D wire at $T=0$ was derived in Ref.~\cite{ThunKink} neglecting phase gradients in superconducting leads. This invalidates the derivation, since the charge conservation law is violated.  The theory of Ref.~\cite{ThunKink} predicts 
an unphysical phase jump in the limit of zero normal-layer thickness. In this limit the SNS junction becomes an ideal uniform superconductor, in which the current  is supported by a constant phase gradient without any phase jump.}




\maketitle



In the recent paper (further called ``Paper'')   Thuneberg \cite{ThunKink} investigated theoretically the current-phase relation of  a ballistic SNS junction in a 1D quantum wire (called one quantum channel in the Paper). Starting from pioneer works \cite{Kulik,Ishii,Bard} the investigations used the method of the self-consistent  field, in which one  should solve the Bogolyubov\,\textendash\,de Gennes equations  together with the self-consistency  equation for the pairing potential. But in the past  the self-consistency  equation was mostly ignored, and instead it was postulated that the superconductor order parameter is constant in superconductors and sharply drops to zero in the normal metal. This approach was used in the Paper under the name ``square-well model''.

\begin{figure}[t]%
\centering
\includegraphics[width=0.5\textwidth]{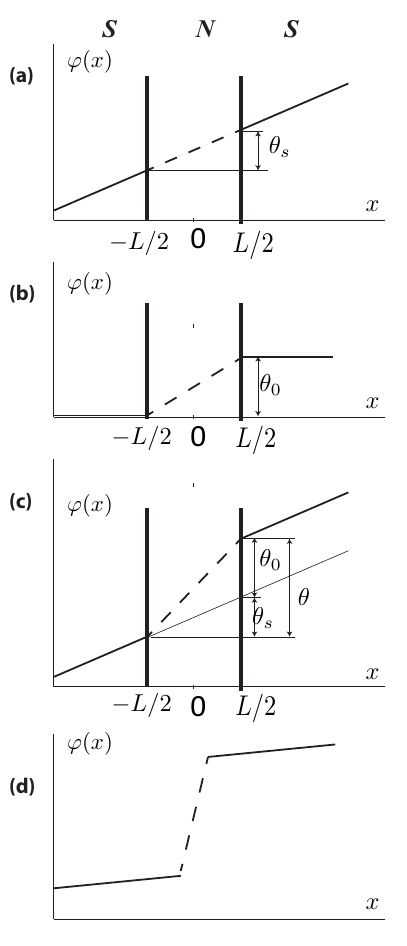}
\caption{The phase variation across the SNS junction.  (a) The condensate current produced by the phase gradient  $\nabla \varphi$ in the superconducting layers.  The phase $\theta_s =L\nabla \varphi$ is the superfluid phase. In all layers the electric current is equal to $env_s$.  (b) The vacuum current produced by the phase $\theta_0$, which is called the vacuum phase. The current is confined to the normal layer, there is no current in superconducting layers.  (c) The  superposition of the condensate  and the vacuum current.  (d) The phase variation across a weak link.}\label{f1}
\end{figure}

In this Comment  I focus  on the $T=0$ case, in which previous investigations predicted for a long ballistic SNS junction the saw-tooth current-phase relation. At $T=0$ there are no thermal quasiparticles.  According to the Bogolyubov\,\textendash\,de Gennes equations, two types of current  states are possible at $T=0$ \cite{Son21,SonAndr}. In the first type of the current state  the order parameter phase $\varphi$ is a linear function of a coordinate in the whole junction as in a uniform superconductor, and the current in all layers is
 \be
 J_s=env_s={ev_f \over L}{\theta_s\over \pi}, 
    \ee{CJ}
    where $n$ is the electron density, $v_s={\hbar\over 2m}\nabla \varphi$ is the superfluid velocity, $v_f$ is the Fermi velocity,    $L$ is the thickness of the normal layer, and  $\theta_s=L\nabla \varphi$ is the phase drop across the normal layer [Fig.~\ref{f1}(a)]. The current $J_s$ and the phase drop $\theta_s$ were called the condensate current and the superfluid phase, respectively  \cite{Son21,SonAndr}.  In the other type of the current state there is a difference between two constant phases in two  superconducting leads [no phase gradients in the leads as shown inFig.~\ref{f1}(b)]. The current $J_v$ in this state was called the vacuum current. It is determined by the phase drop $\theta_0$ called the vacuum phase. Figure~\ref{f1}(c) shows the phase distribution for a mixed state, in which both the condensate and the vacuum currents flow.  The  current phase relation is the relation between the total current $J=J_s+J_v$ and the Josephson phase $\theta=\theta_0+\theta_s$.

While the same condensate current flows in all layers of the junction, the vacuum current flows only in the normal layers in conflict with the charge conservation law. This problem was known in the past, and is connected with broken gauge invariance of the effective Hamiltonian with the pairing potential. So, the solution with the vacuum current is mathematically correct, but is unphysical. Thus, it was argued  \cite{Son21,SonAndr} that the vacuum current is impossible at $T=0$  excepting the phases $2 \pi s+\pi$ ($s$ is an integer), at which the energy of the lowest Andreev level reaches 0 and the level can be occupied. At finite temperatures the quasiparticle current can compensate the vacuum current and restore the charge conservation law.

In the past it was common to ignore  the phase gradients in superconducting leads on the Josephson current. This means that the current-phase relation was derived for the vacuum current.  By default, it was supposed that this is not an issue because gradients (currents) in superconducting leads are so small that their effect on the current in the normal layer is negligible. This is true for a weak link, inside which the phase varies much faster than in the leads as shown in Fig.~\ref{f1}(d).  The long ballistic SNS junction is a weak link only at high temperatures when the current  is very small.  At zero temperature the long SNS junction is not a weak link \cite{Son21}, and the phase gradient in the leads does affect the current in the normal layer. Thus, one should determine   currents in the normal and  superconducting layers  self-consistently.  The current state with the vacuum and condensate current [Fig.~\ref{f1}(c)] is obtained from the state with only the vacuum current [Fig.~\ref{f1}(b)] by the Galilean transformation. The Galilean invariance (despite broken translational invariance) for the case of Andreev scattering was revealed by Bardeen and Johnson \cite{Bard} and was proved in Sec.~3.1 of Ref.~\cite{SonAndr} for an arbitrary superconducting gap profile.  The Galilean transformation cannot restore the charge conservation law because it produces the same condensate current in all layers.  Thus,  the charge conservation law requires that $\theta_0=0$ and there is no vacuum current.

The fact that the current in the ballistic SNS junction is the condensate but not the vacuum current seriously changes the physical picture of the charge transport through the junction. Tuning of the vacuum phase $\theta_0$ determining the vacuum current changes positions of Andreev levels with respect to the gap. This produces a spectral flow, and Andreev levels can enter to or exit from the gap. In contrast, tuning  of the superfluid phase $\theta_s$ determining the condensate current shifts Andreev levels together with the gap without changing their relative positions. Another difference between two current states is different comparative roles of bound and continuum states. The contribution of continuum states to the vacuum current is of the same order as that of the bound states (or even negligible in the multidimensional case or in a short SNS junction). On the other side, the contributions of of bound and continuum states to the condensate current are proportional to their contributions to the density, and the contribution of bound states is negligible in the limit $\Delta_0/\varepsilon_f \to 0$  ($\Delta_0$ is the modulus of the superconducting gap and $\varepsilon_f$ is the Fermi energy).

Despite this essential difference between the vacuum and the condensate current,  the current-phase relations for two currents coincide in the limit $L/\zeta_0=L\Delta_0/\hbar v_f \to \infty$ ($\zeta_0$ is the coherence length), and it was incorrectly concluded that there is no difference between the vacuum and the condensate current (see Refs.~ \cite{Thun,Son23rep}). But this coincidence takes place only in the limit  $L/\zeta_0\to \infty$ \cite{SonAndr}. The Paper addresses the whole diapason of $L$ down to the limit $L=0$. The difference between  the vacuum and the condensate currents becomes very  large  at $L\to 0$. In this limit the  Paper received  the current-phase relation Eq.~(7) derived by Kulik and  Omel'yanchouk \cite{KulOmel,KulOmel2}. But Kulik and  Omel'yanchouk used this equation for narrow bridges between bulk superconducting leads, which are weak links, and the current-phase relation is accurately determined by the vacuum current indeed. This is why the current-phase relation  
\be
J={e \Delta \over \hbar}\sin{\theta\over 2}
     \ee{CJT}
given  by  Eq. (7) at $T=0$ is correct in the case of Kulik and  Omel'yanchouk and is incorrect in the Paper addressing  the SNS junction with all segments of the wire being one-dimensional. This junction is not a weak link, and the correct current-phase relation is given by \eq{CJ} keeping in mind that the condensate current is the only current($J=J_s $, $\theta=\theta_s$).  According to \eq{CJT}, in the limit $L=0$ there is a finite phase difference across the normal layer of zero thickness, i.e, a phase jump. Meanwhile,  in the limit $L=0$ the 1D SNS junction becomes a uniform superconductor. The phase jump in the limit $L=0$  is a challenge  to common sense.  How is it possible that the normal layer dramatically affects the phase distribution in the current state even after it was removed from the wire? On the other hand, the limit $L=0$ for a narrow bridge between two bulk superconductors addressed by Kulik and  Omel'yanchouk reduces to a fluid flowing through a small hole in a thin partition wall, and the phase jump is quite natural.

 Because of Galilean invariance, the expression \eq{CJ} for the  condensate current  is valid  for an arbitrary superconducting gap profile and for any thickness $L$ of the normal layer.   It is exact in the limit $\Delta_0/\varepsilon_f \to 0$. Using the condensate current for the current-phase relation instead of the vacuum current eliminates the problem with the charge conservation law within the square-well model. Therefore, the statement of the Paper that ``attempts to restore current conservation without solving the self-consistency of the pair potential are misguided'' is wrong. This is fully confirmed by numerical calculations of  the Bogolyubov\,\textendash\,de Gennes equations together with the integral self-consistency equation for the gap by Riedel {\em et al} \cite{Bagwell}. They revealed that although the absolute value of the gap smoothly varies between the normal layer and the  superconducting leads, the phase gradient is constant in all layers  as expected for the condensate current [Fig.~\ref{f1}(a)].  
 
Although the vacuum current given by Eq.~(7) cannot flow at $T=0$,  it can flow at finite temperatures when its violation of the charge conservation law is compensated by the current produced by quasiparticles. Here 
I cannot help but comment the theoretical formalism used in the Paper for its derivation. Equation (3) is determined as the ``starting point'' of the derivation. This equation is an expression for the current via a sum over Matsubara poles. The derivation of this expression by Ishii \cite{Ishii} using the temperature Green function formalism
 occupies about 10 pages and is described by the Paper itself as ``rather complicated''. Then the Paper transforms Eq.~(3) to the expression Eq.~(5), which contains a sum over bound states and an integral over continuum states and is in fact the {\em ab initio} expression for the current, from which Ishii must start. The expression directly follows from the scattering theory for the Bogolyubov\,\textendash\,de Gennes equations \cite{Bard,Son21,Bagw}. One may ask whether using of the sophisticated temperature Green function formalism was needed.

In summary, the current-phase relation  for the 1D ballistic SNS junction at $T=0$ calculated in the Paper is in conflict with the charge conservation law and is incorrect.
 


\end{document}